\documentstyle[preprint,prb,aps]{revtex}
\large
\begin{document}
\newcommand{\be}{\begin{equation}}
\newcommand{\ee}{\end{equation}}
\newcommand{\bea}{\begin{eqnarray}}
\newcommand{\eea}{\end{eqnarray}}
\draft

\title{ Anderson-Yuval approach to the multichannel Kondo problem}
\author{
M. Fabrizio$^{(a,b)}$, Alexander O. Gogolin$^{(a,c)}$,
and Ph. Nozi\`eres$^{(a)}$ }
\address{
$^{(a)}$Institut Laue-Langevin, B.P.156 38042 Grenoble, Cedex 9, France\\
$^{(b)}$Institut for Advanced Studies, Via Beirut 4, 34014 Trieste, Italy\\
$^{(c)}$Landau Institute for Theoretical Physics, Kosygina str. 2,
Moscow, 117940 Russia
}
\date{\today}
\maketitle
\begin{abstract}
We analyze the structure of the perturbation expansion of
the general multichannel Kondo model with channel anisotropic exchange
couplings
and in the presence of an external magnetic field,
generalizing to this case the Anderson-Yuval technique.
For two channels, we are able to map the Kondo model onto a generalized
resonant level model. Limiting cases in which the
equivalent resonant level model is solvable are identified.
The solution correctly captures the properties of the two channel
Kondo model, and also allows an analytic description of the cross-over from
the non Fermi liquid to the Fermi liquid behavior caused by the channel
anisotropy.
\end{abstract}
\narrowtext

\section{Introduction}
\label{sec:Introduction}

The single channel Kondo model has a long history
as the simplest model believed to contain the relevant physics
of magnetic impurities
embedded in metals. A lot of efforts have been devoted to
study this model and presently one can safely claim that
it has been completely understood from the theoretical point of
view. Apart from the original perturbative scaling approach,
which already gave the correct qualitative description
\cite{AY&H,singleKondo}, there is also an exact solution of
this model obtained by Bethe-ansatz technique\cite{Betheansatz-single}.
The physics underlying the single channel Kondo model is
the formation of a non degenerate singlet state at low
temperature. The impurity spin is screened by the conduction electrons,
hence the magnetic susceptibility,
obeying Curie-Weiss law at high temperatures, undergoes the
Kondo cross-over saturating to a constant upon lowering the temperature.

In an attempt to describe realistic magnetic impurities
(which have orbital structure), different generalizations
of the simple Kondo model have been proposed\cite{generalizations}.
Among these generalizations, the simplest one is probably
the model describing an impurity spin $S$ coupled to
$N$ channels of conduction electrons (commonly referred to as
the multichannel Kondo model). Surprisingly, it has been realised
that, for $N>2S$, this model exhibits a behavior
qualitatively different from the single channel one\cite{N&B},
which is due to a non-trivial ground state with a residual degeneracy.
This gives rise to divergent low temperature susceptibility $\chi_{imp}$
and specific heat coefficient $\gamma=C_{V, imp}/T$
(so-called non Fermi liquid behavior).
It should be said that the experiments on dilute magnetic
alloys (the systems, for which the Kondo model was
originally proposed) do not give any clear evidence for such a behavior
but rather suggest that the ground state is always a singlet.
A possible explanation would be that the channel symmetry
(which is necessary for the non Fermi liquid behavior to occur)
is broken (since typically no exact symmetry guarantees channel
equivalence). If the energy scale of this symmetry breaking
term is small, a slow cross-over from the non Fermi liquid to the
Fermi liquid behavior is expected. However, it might be difficult
to experimentally distinguish it from the usual Kondo cross-over.
Consequently, since early 80's, the interest to the issue
was decaying, even though,
in the meanwhile, some exact solutions
for the case of equivalent channels became available.

In recent years, however, interest to the subject grew again,
as other, more promising, realizations of the multichannel
Kondo model have been proposed; for instance, two level systems in
metals\cite{TLS}, heavy fermion compounds\cite{HFC},
high Tc superconductors\cite{HTcS}. While for the two level systems
there are convincing experimental evidences\cite{Ralph}
of the non Fermi liquid behavior predicted by the theory, the other
proposed realizations are till now controversial.

The general multichannel Kondo hamiltonian is given by
\begin{equation}
H_K= \sum_{a=1}^N \sum_\sigma
H_0(\psi^{\phantom{\dagger}}_{a\sigma},\psi^\dagger_{a\sigma})
+\sum_{a=1}^N \left\{ J_{z a} S^z \sigma_a^z({\bf 0})
+ \frac{J_{\perp a}}{2} \left[ S^+ \sigma_a^-({\bf 0})
+ S^- \sigma_a^+({\bf 0}) \right] \right\},
\label{ham-Kondo}
\end{equation}
where
\be
H_0(\psi,\psi^\dagger)=\sum_{{\bf k}}
\epsilon_{{\bf k}} \psi^\dagger_{{\bf k}}
\psi^{\phantom{\dagger}}_{{\bf k}}
\label{H0}
\ee
is the kinetic energy of the conduction electrons $\psi_{a\sigma}$,
$a=1,...,N$ being the channel index, and
$\sigma=\uparrow,\downarrow$ being the spin index (we assume
a spin 1/2 impurity).
Notice however that in some realizations
of this model, the channel index is the physical spin while the
spin index labels an orbital quantum number (see also Section IV).
The electron spin densities in Eq.~(\ref{ham-Kondo}) are
defined by
\be
\begin{array}{lcl}
\displaystyle
\sigma^z_{a}({\bf x}) &=& \frac{1}{2}\left[ \psi_{a\uparrow}^\dagger({\bf x})
\psi_{a\uparrow}^{\phantom{\dagger}}({\bf x})
- \psi_{a\downarrow}^\dagger({\bf x})
\psi_{a\downarrow}^{\phantom{\dagger}}({\bf x})\right]\\
\displaystyle
\sigma^+_{a}({\bf x}) &=& \psi_{a\uparrow}^\dagger({\bf x})
\psi_{a\downarrow}^{\phantom{\dagger}}({\bf x}).\\
\end{array}
\label{spindensity}
\ee

In the single channel case ($N=1$)
the exchange couplings $J$ flow to infinity under scaling, if positive.
The spin anisotropy $J_z\neq J_\perp$ disappears at the fixed point.
This is interpreted as the formation of a singlet at the
impurity site. On the contrary for the $N>1$ {\em channel symmetric}
case, the infinite coupling fixed point is unstable as well as the weak
coupling one,
hence a stable intermediate coupling fixed point.

Until recently, the information on the
behavior of the model around this fixed point could only be
extracted from
the Bethe-ansatz solution\cite{Betheansatz} and conformal field theory
\cite{A&L}. In 1992, Emery and Kivelson\cite{E&K}
provided a simple solution for the symmetric two channel case at a
particular value of the longitudinal exchange coupling.
The solution was obtained by means of abelian bosonization technique
generalizing a procedure previously applied
to the single channel Kondo model by Schlottmann\cite{Schlottmann}.
The original Schlottmann's approach was in turn inspired
by Toulouse's mapping of the single channel Kondo
model onto a resonant level model, which he achieved by the analysis
of the partition function\cite{Toulouse}, borrowing
the perturbative treatment of Anderson and Yuval\cite{A&Y}.
The two methods, bosonization and Anderson-Yuval method, are equivalent,
in the sense that they give qualitatively similar results
(it has been checked in the single channel case).
Still, the latter is formally more rigorous (e.g. it does not
fully relies on band linearization) and straightforward
(it simply amounts to compare the perturbation expansions
of two different models).

In this paper we generalize the Anderson-Yuval method to
the multichannel Kondo model with channel anisotropic exchange
couplings. For the two channel case,
we are able to show that the perturbation expansion of the Kondo
model is equivalent to that of a generalized resonant level model.
In the channel isotropic case, the model is of the type found
by Emery and Kivelson via bosonization technique.
The novel feature is the channel anisotropy
which gives rise to interesting cross-over phenomena.
We demonstrate that
even in this case a mapping of the Kondo model (for particular
values of the longitudinal exchange couplings)
onto a solvable resonant level model does exist.
For $N>2$ we are unable to find any kind of resonant level model
which would reproduce the perturbation expansion.

\section{Generalization of the Anderson Yuval approach}
\label{sec:Generalization}

In this Section, we analyze the structure of the perturbation
expansion of the hamiltonian (\ref{ham-Kondo}) in the transverse
exchange couplings.

\subsection{Single channel model}

Consider first the single channel problem. We allow for an exchange anisotropy
($J_{x} = J_{y}= J_\perp$ is different from $J_{z}$). We want to calculate the
impurity partition function in {\em time} space, using a perturbation expansion
in powers of $J_\perp$. A term of order $2n$ involves $2n$ alternate impurity
spin flips. Let $t_{i}$ be the times of up flips, $t'_{i}$ that of down flips
($n$ of each).
The philosophy is to calculate that particular term exactly, for a given
history $\{t_{i},t'_{i}\}$, and to show that it is identical to the
corresponding
expansion for another problem (a resonant level), with
appropriately chosen parameters.
The two problems are mapped on each other term by term:  they are equivalent.
Note that we thus bypass summation of the perturbation series in $J_\perp$. For
a specific value of $J_{z}$ (the "Toulouse limit") the equivalent problem
happens to be trivially solvable: if we can scale through that value we have an
explicit description of the crossover to low temperature - the "($100-
\epsilon$)\%" exact solution of Anderson. The error stems from the fact that
universal scaling is not just a change of $J_{z}$. That error is supposed not
to change the qualitative behavior - and anyway it is implicit in the
equivalent bosonization technique (less powerful since it relies on a Born
approximation for phase shifts).

Assume first that $J_{z}=0$.
Each vertex flips a conduction electron spin.
A $t_{i}$ vertex creates a $\downarrow$ electron and an $\uparrow$ hole,
a $t'_{i}$ vertex does the reverse.
In a typical diagram, the electron propagators go
from any $t_{i}$ to any $t'_{i}$
for $\downarrow$ spins, and from any $t'_{i}$ to any $t_{i}$
for $\uparrow$ spins.
Since $J_{z} =0$ these propagators are {\em free electron local} propagators
\[
G_0(t)=\frac{i \nu_{0}}{t-i\xi_0^{-1}{\rm sign}t},
\]
where $\nu_o$ is the density of states for one spin at
the Fermi level, and $\xi_0$ is a high energy cut-off of the order of
the conduction bandwidth.
Let $D_\sigma(t_{i},t'_{ i})$ be the contribution of the $n$ lines
with spin $\sigma$ that join $t_{i}$ and $t'_{i}$ vertices.  $D_\sigma$ has a
pole whenever $t_{ i}=t'_{ j}$ (one propagator is divergent). Moreover {\em
crossing symmetry} implies a zero whenever $t_{i}=t_{j}$ (or $t'_{i}=t'_{j}$):
exchanging the extremities of two propagators changes the sign.
Hence the Cauchy determinant found by Anderson and Yuval
(here and in what follows we omit time independent pre-factors containing the
density of states $\nu_0$ and to the short-time cut-off, restoring them
in the final expressions)
\be
D_\sigma=\frac{\displaystyle
\prod_{i<j}(t_i-t_j)\prod_{i<j}(t'_i-t'_j)}
{\displaystyle
\prod_{i,j}(t_i-t'_j)}\; .
\label{CauchyAnd}
\ee
Expression (\ref{CauchyAnd}) is homogeneous
with degree $(-n)$, as expected for $D_\sigma$.
The proof is completed by looking at the asymptotic behavior.
The corresponding contribution to the impurity propagator
is $U_0=D_\uparrow D_\downarrow$.

We now restore $J_{z}$. The potential felt by a spin $\sigma$ electron changes
at each impurity flip -- hence an edge singularity that will modify the long
time behavior of $U$, see Ref.\onlinecite{N&D}.
In Fig.1 we have drawn the time dependent potential
felt e.g. by the up-spin conduction electrons at the impurity site
(the down-spin one is the opposite)
The phase shift is $\delta_{+}$ (resp.  $\delta_{-}$) when the
electron and impurity have parallel (resp.  antiparallel) spins.
What matters is the {\em discontinuity} of phase shifts when a flip occurs,
$\delta=\delta_{+}-\delta_{-}$. If we assume electron hole symmetry,
then
\be
\delta_{+}=-\delta_{-}=\tan^{-1}(\pi\nu_0J_{z}/4)
\label{delta}
\ee

The effect of the flip is two-fold:
\begin{itemize}
\item[({\em i})] The {\em open lines} that contribute to $D_\sigma$ can scatter
any number of times
on the impurity.  That generates once again the
Cauchy determinant (via a Muskhelishvili type of analysis\cite{Mush}).
For each spin $D_\sigma$ is replaced by $D^{1-2\delta/\pi}$, thus
open line contribution acquires an extra factor $U_L=D^{-4\delta/\pi}$.
\item[({\em ii})] In addition the underlying Fermi sea reacts to the flip via
{\em closed loops} that exponentiate (for a given history the potential is
structureless). The resulting contribution is $U_C=D^{2\delta^2/\pi^2}$.
\item[({\em iii})] Altogether $U=U_0U_LU_C$, and, after
inserting back the prefactors, we obtain
\begin{equation}
U=\left(\frac{J_\perp\nu_0\xi_0}{2}\right)^{2n}
\left( \frac{D^\eta}{\xi_0^n}\right)^\eta,
\label{U-single}
\ee
where $\eta$ is an exponent that
depends on $J_z$,
\[
\eta=2\left(1-\frac{\delta}{\pi}\right)^2.
\]
(Note that $\eta=0$ if $\delta=\pi$, which corresponds to the strong
coupling limit $\delta_{+}=-\delta_{-}=\pi/2)$.
\end{itemize}

Let us now consider a resonant level model for spinless electrons,
characterized by the hamiltonian
\be
H=H_0(\Psi,\Psi^\dagger) +
\lambda \left[
\Psi^\dagger({\bf 0})d + d^\dagger\Psi({\bf 0})\right] +
\frac{V}{2}\left[
\Psi^\dagger({\bf 0})\Psi({\bf 0}) -
\Psi({\bf 0})\Psi^\dagger({\bf 0}) \right]
\left( d^\dagger d -\frac{1}{2} \right)
\label{reslevel}
\ee
where $d$ is an impurity orbital at the Fermi energy located at the origin.
$\Psi$ is a free Fermi field which kinetic energy is the same as in
Eq.(\ref{ham-Kondo}).
The interaction potential $V$ produces a phase shift discontinuity
\begin{equation}
\delta'=2{\rm tan}^{-1}\left(\frac{\pi\nu_0 V}{2}\right)
\label{deltaprime}
\end{equation}
between the empty and full d-states.
We expand in powers of $\lambda$ which plays the role of $J_\perp$.
The $t_{i}$ and $t'_{i}$ vertices correspond to $d^\dagger$ and $d$ operators.
The structure of the expansion is the same as for the Kondo case, except that
there is no spin degeneracy.
We have one set of open lines originating from $\lambda$ vertices,
which can scatter off the flipping $\delta'$.
We also have one set of closed loops -- hence altogether
\be
U'=\left( \lambda \sqrt{\nu_0\xi_0}\right)^{2n}
\left(\frac{D}{\xi_0^n}\right)^{\eta '}
\label{simple-resonant}
\ee
with
\[
\eta'=\left(1-\frac{\delta'}{\pi}\right)^2.
\]

Thus the Kondo problem with coupling $\delta$ is mapped term by term
onto the resonant level with coupling $\delta'$ if the two propagators
Eqs~.(\ref{U-single}) and (\ref{simple-resonant}) are identical. This
implies $\eta=\eta'$ (that can always be achieved by appropriately
choosing $V$) and
\[
\lambda= \frac{J_\perp}{2}\sqrt{\nu_0\xi_0}.
\]
The Toulouse limit corresponds to $\delta'=0$, i.e.  a phase shift
$\delta=\pi(1-1/\sqrt{2})$. The resonant level
hamiltonian can then be trivially diagonalized, yielding a low
temperature Fermi liquid behavior. The resonant level hamiltonian
of the form (\ref{reslevel}) has been previously derived by Wiegmann
and Finkelstien \cite{Finkelstein}.

\subsection{Multichannel model}

We now turn to the N channel case.  In a first stage we assume flavor
degeneracy:
what does remain of the previous analysis?  In order to answer that
question we proceed in reverse.
\begin{itemize}
\item[({\em i})] The alternation of up and down spins is unchanged.
The flipping potential due to $\delta$ is the same as before
(see Fig.1), whatever the flavor
involved at each vertex.
Edge singularities are consequently unaffected.  The scattering
contribution to open lines is again $U_L=D^{-4\delta/\pi}$
(flavor is fixed by extremities).
Each closed loop can have an arbitrary flavor and therefore
$U_C=D^{2N\delta^2/\pi^2}$.
\item[({\em ii})] Paradoxically the difficulties come from the part $U_{0}$
(in the absence of $J_{z}$).  $U_{0}$ still has poles whenever $t_{i}=t'_{j}$,
but one {\em looses crossing symmetry}.
If the ends of two lines are interchanged
one usually changes the number of closed loops C -
hence a change in the degeneracy $N^{C}$.
As a result $U_{0}$ cannot be expressed simply in terms of $D$.
\end{itemize}

In order to proceed, we must assign to each vertex its flavour index $a$.
The $t_{i}$ and $t'_{j}$ then break into $N$ subclasses, $t_{ia}$ and $t'_{ja}$
($a=1,...,N$).
For a given diagram the number of vertices in different subclasses needs not
be the same,
but spin and flavor conservation implies an equal number of $t_{ia}$ and
$t'_{ja}$ within a given subclass.  Since flavor is
conserved along an open line, $U_{0}$ is a product of independent factors
$U_{0a}$.
For each factor crossing symmetry holds and $U_{0a}$ is the square
of a Cauchy determinant $D_a$ as it would be for a single channel.
$D_a$ is still given by (\ref{CauchyAnd}), the products
running only over $a$-type times, $t_{ia}$ and $t'_{ja}$.
In the end we find (again omitting prefactors proportional to
$J_\perp$, $\nu_0$ and $\xi_0$)
\be
U=\left[ D_1 ... D_N \right]^2 D^{-4\frac{\delta}{\pi}+
2N \left[\frac{\delta}{\pi}\right]^2}=
\frac{\left[ D_1 ... D_N \right]^2}{D}D^{\beta_N}
\label{Umulty}
\ee
where
\be
\beta_N=1-4\frac{\delta}{\pi}+2N\left(\frac{\delta}{\pi}\right)^2.
\label{beta}
\ee
Note that $D$ is {\em not} the product of individual $D_a$:
we have instead $D= D_1 ... D_N F$, in which we have set
\[
F=\frac{\displaystyle
\prod_{ij}\prod_{a<b}
\left( t_{ia} - t_{jb} \right) \left( t'_{ia} - t'_{jb} \right)}
{\displaystyle
\prod_{ij}\prod_{a<b}
\left( t_{ia} - t'_{jb} \right) \left( t_{ia} - t'_{jb} \right)}
\]
The factor $F$ couples the channels.

The Emery-Kivelson solution to the two channel case, $N=2$,
is based on a mapping of the Kondo problem onto the following
spinless resonant level hamiltonian
\bea
H&=&H_0(\Psi,\Psi^\dagger) + H_0(\Psi_s^{\phantom{\dagger}},\Psi_s^\dagger)
+ \lambda \left(d^\dagger - d \right) \left[
\Psi^\dagger({\bf 0}) + \Psi({\bf 0})\right] \nonumber\\
&+& \frac{V}{2}\left[
\Psi_s^\dagger({\bf 0})\Psi_s^{\phantom{\dagger}}({\bf 0})
-\Psi_s^{\phantom{\dagger}}({\bf 0})\Psi_s^\dagger({\bf 0})\right]
\left( d^\dagger d -\frac{1}{2} \right)
\label{EKham}
\eea
where $d$ is again a fictitious spinless Fermi operator.
Notice that we have introduced two Fermi fields
$\Psi$ and $\Psi_s$, coupled to the impurity in a different way
(the reason why we have not used the same field will become clear
later).
In order to establish the equivalence we first consider the case $V=0$.
We divide the $d$ and $d^\dagger$ flips into two subclasses,
depending on whether a fermion $\Psi$ is emitted or absorbed.
Times $t_{i1}$ and $t_{i2}$ correspond to $d^\dagger$ flips with a fermion
emitted or absorbed respectively,
$t'_{i1}$ and $t'_{i2}$ are their hermitian conjugates.
A spinless fermion propagator can go as usual from $t_{i1}$ to $t'_{j1}$
($t'_{i2}$ to $t_{j2}$) {\em or}
from $t_{i1}$ to $t_{j2}$ ($t'_{i2}$ to $t'_{j1}$).
The latter possibility is the new feature.
The corresponding impurity propagator $U'$ has poles whenever
a propagator has zero time range, i.e. when $t_{ia}=t'_{ja}$,
$t_{i1}=t_{j2 }$ or $t'_{i2}=t'_{j1}$.
Due to crossing symmetry it has zeroes when
$t_{ia}=t_{ja}$, $t'_{ia}=t'_{ja}$,  $t_{i1}=t'_{j2}$ or
$t'_{i1}=t_{j2}$. Once again one thereby builds a Cauchy determinant
which happens to be
\be
U'=\frac{D_1 D_2}{F}
\label{Ustar}
\ee
Expression (\ref{Ustar}) has the right poles and zeroes.
It moreover has the right overall power of $t$ and asymptotic behavior:
it is the correct answer.
Comparing (\ref{Ustar}) with the definition of $F$ we see that
\[
U'=\frac{\left(D_1 D_2\right)^2}{D}
\]
We now restore the flipping potential $V$.
Since it involves a different Fermi field, it gives rise only to
a closed loop contribution.
Altogether we have
\begin{equation}
U'=\frac{\left(D_1 D_2\right)^2}{D}
D^{\left(\frac{\delta'}{\pi}\right)^2}
\label{Ustarprime}
\end{equation}
with the same $\delta'$ as in Eq.~(\ref{deltaprime}).
Comparing (\ref{Ustarprime}) with (\ref{Umulty}) we see that
the two problems are mapped onto each other if [see Eq.~(\ref{beta})]
\be
\beta_2=\left(1-2\frac{\delta}{\pi}\right)^2=
\left(\frac{\delta'}{\pi}\right)^2
\label{xi}
\ee
which is always possible since both right and left sides are positive.
Notice that for a given $\delta$ the interaction potential $V$ in
Eq.~(\ref{EKham}) is, according to (\ref{deltaprime})
and (\ref{xi}), given by
\be
V=\frac{2}{\pi\nu_0}\tan\left(\frac{\delta'}{2}\right)
=\frac{2}{\pi\nu_0}\tan\left(\frac{\pi}{2}-\delta \right)
\label{V}
\ee

The problem is directly solvable if $V=0$, i.e. when $\delta=\pi/2$.
In the electron hole symmetric case that implies $\delta_{+}=-\delta_{-}=
\pi/4$ -- a typical intermediate coupling as expected for the two channel
overscreened Kondo impurity.
Indeed from the expression of $\beta_2$ we see that the model is symmetric
under $\delta\to \pi - \delta$. This extends the result of
Ref.\onlinecite{N&B}
that the two channel Kondo model behaves similarly around $J_z=0$
(i.e. $\delta=0$) and $J_z=\infty$ (i.e. $\delta=\pi$). By symmetry
the fixed point should be exactly at $\delta=\pi/2$, that is at the
solvable line $V=0$.

The argument can be extended to a flavor dependent exchange $J$.
Due to anisotropy we must treat separately the channel dependence of
$J_\perp$ and $J_{z}$.
Different $J_{\perp 1}$ and $J_{\perp 2}$ do not affect the structure of the
perturbation expansion. As we have shown above the mapping works as follows:
\[
\begin{array}{lcr}
\Psi_{1\downarrow}^\dagger({\bf 0},t_{i1})
\Psi_{1\uparrow}^{\phantom{\dagger}}({\bf 0},t_{i1}) S^+(t_{i1}) &
\longmapsto & d^\dagger(t_{i1}) \Psi({\bf 0},t_{i1});\\
\Psi_{2\downarrow}^\dagger({\bf 0},t_{i2})
\Psi_{2\uparrow}^{\phantom{\dagger}}({\bf 0},t_{i2}) S^+(t_{i2}) &
\longmapsto & d^\dagger(t_{i2}) \Psi^\dagger({\bf 0},t_{i2});\\
\Psi_{1\uparrow}^\dagger({\bf 0},t'_{i1})
\Psi_{1\downarrow}^{\phantom{\dagger}}({\bf 0},t'_{i1}) S^-(t'_{i1}) &
\longmapsto & \Psi^\dagger({\bf 0},t'_{i1}) d(t'_{i1});\\
\Psi_{2\uparrow}^\dagger({\bf 0},t'_{i2})
\Psi_{2\downarrow}^{\phantom{\dagger}}({\bf 0},t'_{i2}) S^-(t'_{i2}) &
\longmapsto & \Psi({\bf 0},t'_{i2}) d(t'_{i2}),\\
\end{array}
\]
thus we need only modify accordingly the flipping matrix elements of the
equivalent model, which becomes
\be
\frac{J_{\perp 1}}{2}\sqrt{\nu_0\xi_0}
\left[d^\dagger \Psi({\bf 0}) + \Psi^\dagger({\bf 0}) d\right]
+\frac{J_{\perp 2}}{2}\sqrt{\nu_0\xi_0}
\left[d^\dagger \Psi^\dagger({\bf 0}) + \Psi({\bf 0}) d\right]
\label{matel}
\ee
Notice that, if $J_{\perp 1}=J_{\perp 2}$, (\ref{matel})
reduces to (\ref{reslevel})
with $\lambda=J_{\perp 1}\sqrt{\nu_0\xi_0}$.

A difference between $J_{z1}$ and $J_{z2}$
gives rise to different phase shifts $\delta_{1}$ and $\delta_{2}$.
Let us first consider the scattering correction to the $"1"$ open line,
$U_{L1}$.
The Muskhelishvili propagator for a channel $1$ spin up electron is
\[
G_1(t,t')=\frac{i\pi\nu_0}{t-t'}\prod_{i,a}
\left[
\frac{\left(t-t_{ia}\right)\left(t'-t'_{ia}\right)}
{\left(t-t'_{ia}\right)\left(t'-t_{ia}\right)}
\right]^{-\frac{\delta_1}{\pi}}
\]

Its contribution to $U_{L1}$ is
obtained by putting $t$ equal to any $t_{i1}$, $t'$  to any $t'_{i1}$,
hence a factor $\left[D_1^2F\right]^{-\delta_1/\pi}$.
We square it in order to account for spin and we multiply by the corresponding
term for
channel $2$.
The closed line contribution is straightforward since flavor is conserved along
a loop,
\[
U_C=D^{2\left(\delta_1^2+\delta_2^2\right)/\pi^2}
\]
Altogether the impurity propagator is
\be
U=D_1^{2-4\delta_1/\pi} D_2^{2-4\delta_2/\pi}
F^{-2\left(\delta_1+\delta_2\right)/\pi}
D^{2\left(\delta_1^2+\delta_2^2\right)/\pi^2}
\label{againU}
\ee
(Remember that $F=D/D_{1}D_{2}$.) If $\delta_{1}=\delta_{2}=\delta$,
we recover the previous result (\ref{Umulty}).

In general let us write
\[
\delta_{1}=\delta + \varepsilon \;,\;\;\delta_{2}=\delta - \varepsilon
\]
Then (\ref{againU}) reduces to
\be
U=\frac{\left(D_1 D_2\right)^2}{D}D^{\beta_2}
D^{4\left(\frac{\varepsilon}{\pi}\right)^2}
\left(\frac{D_2}{D_1}\right)^{4\frac{\varepsilon}{\pi}}
\label{againUbis}
\ee
The additional factors with respect to
(\ref{Ustarprime}) can be reproduced with an extra potential
\[
\frac{W}{2}\left[
\Psi^\dagger({\bf 0})\Psi({\bf 0})
-\Psi({\bf 0})\Psi^\dagger({\bf 0})\right]
\left(d^\dagger d - \frac{1}{2}\right)
\]
With the choice
\be
W=\frac{2}{\pi\nu_0}\tan\varepsilon
=\frac{2}{\pi\nu_0}\tan\left(\frac{\delta_1-\delta_2}{2}\right)
\label{W}
\ee

the closed loop contribution generates the factor
$D^{\left(2\varepsilon/\pi\right)^2}$.
The last factor of (\ref{againUbis}) comes from the fact that the phase shift
discontinuity is $+\varepsilon$ on the $"1"$ vertices, $-\varepsilon$ on
the $"2"$ vertices (see Fig.~2).
In that way one can map any version of the two channel Kondo impurity onto an
extended Emery-Kivelson hamiltonian.

It is interesting to examine to what
extent such an analysis could be pursued if  $N>2$.
We return to the flavor symmetric case,
for which (\ref{reslevel}) holds.
If we manage to have $\beta_N=0$, then
\be
U=\frac{\left(D_1 ... D_N\right)^2}{D}=\frac{D_1 ... D_N}{F}
\label{Ubisbis}
\ee
Hence two questions:  ({\em i}) Can we achieve $\beta=0$?  ({\em ii}) If we
can,
is there a solvable model that gives the same $U$?
It is clear that no real phase shift $\delta$ will achieve $\beta=0$ if $N>2$.
That may be a definitive objection since poor man's scaling scans the real
$\delta$
axis.  Let us ignore it, hoping that some analytic continuation argument might
help.
Then in order to reproduce (\ref{Ubisbis}) we must introduce a coupling
\[
S^+\left(\Psi_1+...+\Psi_N\right)+{\rm H.c.}
\]
in which the $\Psi_{a}$ operators are such that the corresponding propagators
are
\be
\begin{array}{rcl}
\langle  \Psi^{\phantom{\dagger}}_a (t) \Psi^\dagger_b (t')\rangle & = &
g(t-t') \left( 1- \delta_{ab} \right) \\
\langle \Psi^{\phantom{\dagger}}_a (t)
\Psi^{\phantom{\dagger}}_b (t')\rangle & = & g(t-t') \delta_{ab}\\
\end{array}
\label{algebra}
\ee
($g(t) \approx 1/t$ is the free electron propagator).
Then $U$ will have poles whenever $t_{ia}=t'_{ja}$ on the one hand,
$t_{ia}=t_{jb}$, $t'_{ia}=t'_{jb}$ ($a \neq b$) on the other.
It will have zeroes if $t_{ia}=t_{ja}$, $t'_{ia}=t'_{ja}$ or $t_{ia}=t'_{jb}$.
That just generates the combination (\ref{Ubisbis}).
It remains to be seen what kind of algebra could
produce (\ref{algebra}):  we do not know of any.

\section{Magnetic field effects}

Let us consider the effects of an uniform  magnetic field
$\vec{B}=(0,0,B)$
in the framework of the Anderson-Yuval approach. The magnetic
field appears in the hamiltonian with a term
\be
H_B=-\mu_B B\left[g_{i} S^z + g_c \int d{\bf x}
\sum_{a=1}^N \sigma_{a z}({\bf x}) \right]
\label{magnetic}
\ee
where the electron spin density is defined in Eq.~(\ref{spindensity}),
$g_{i}$ and $g_c$ are the Land\'e factors of
the impurity and the conduction electrons respectively,
and $\mu_B$ is the Bohr magneton.

As before we will treat
the transverse exchange perturbatively. This implies
that the reference states $\mid \uparrow \rangle$
and $\mid \downarrow \rangle$, which are used for the perturbation
expansion, are the eigenstates of the hamiltonian with
fixed impurity spin direction in the presence of the magnetic field
\be
H_{\uparrow/\downarrow}=\sum_{a \sigma} H_0(\psi_{a\sigma}^\dagger,
\psi_{a\sigma}^{\phantom{\dagger}})
\pm \sum_{a=1}^N \frac{J_{az}}{2} \sigma_{a z}({\bf 0})
\mp \frac{g_i \mu_B}{2} B - g_c \mu_B B \sum_{a=1}^N
\int d{\bf x} \sigma_{az}({\bf x}).
\label{ham-magnetic}
\ee

We have to understand how the magnetic field modifies the perturbation
expansion
in $J_\perp$. $B$ gives rise to two effects.
\begin{itemize}
\item[({\em i})] It shifts the chemical potential for up and
down spin electrons (in opposite directions).
This causes a small change in the spin up and down phase shifts if the band
has a finite curvature at the Fermi energy. This effect is negligible
at low temperature.
\item[({\em ii})] It causes a difference $\Delta E=E_\uparrow -
E_\downarrow$ in the ground state energies
of (\ref{ham-magnetic}) for the two impurity spin directions, which
appears in the Muskhelishvili propagators.
\end{itemize}

By standard phase shift arguments, based on Friedel's sum rule for the
displaced charge,
we find
\be
\Delta E = -\mu_B g_i B + \frac{1}{\pi}
\sum_{a=1}^N
\int_{\epsilon_F - \frac{g_c \mu_B}{2} B}
^{\epsilon_F + \frac{g_c \mu_B}{2} B}  {
d\epsilon \delta_{a}(\epsilon) }
\label{DeltaE}
\ee
where $\epsilon_F$ is the Fermi energy. For small magnetic field
(\ref{DeltaE}) reduces to
\be
\Delta E = -\mu_B g_i B + \frac{g_c\mu_B }{\pi} B
\sum_{a=1}^N \delta_{a}.
\label{DeltaEbis}
\ee
The above energy difference enters in the impurity propagator
(\ref{Umulty}) via the following phase factor
\be
\exp{\left[-i\Delta E\sum_{i=1}^n (t_i - t'_i)\right]}.
\label{phase}
\ee
The conduction electron part of $\Delta E$ actually represents
the leading term of closed loops diagrams, that one which grows linearly with
$(t_i-t'_{i})$ instead of logarithmically.

Which term has to be added
to the resonant level model in order to reproduce (\ref{phase})?
It is easy to realize that the corresponding term is simply
\be
\Delta E S_z \longmapsto \Delta E \left(d^\dagger d -\frac{1}{2}\right).
\ee

Notice that at the Emery-Kivelson line for the two channel case
\[
\sum_{a=1,2} \delta_{a} = \pi
\]
so that if $g_i=g_c$ then $\Delta E=0$ (at first order in $B$).
Consequently the impurity magnetic susceptibility vanishes,
in agreement with conformal field theory\cite{A&L} and bosonization
approaches\cite{others}.
This in turns means that at the Emery-Kivelson line the
reference states are such as to perfectly screen the impurity
spin.
When the departure away from the Emery-Kivelson line is treated as a
perturbation\cite{others}, both the specific heat and susceptibility
acquire logarithmic sigularities, leading to the universal
Wilson ration $R_W=8/3$.

As to the $N>2$ channel symmetric case,
conformal field theory\cite{A&L} and
abelian bosonization approaches (which till now exist only for
$N=4$, see Ref.~\onlinecite{fourchannel}) again predict
the impurity susceptibility to vanish at the fixed point.
{}From Eq.~(\ref{DeltaEbis}), we see that $\Delta E=0$
for $\delta=\pi/N$, and therefore the impurity susceptibility
is rigorously zero. If this is the true property of the fixed point,
as it follows from the analysis of Ref.~\onlinecite{A&L}, then
$\delta=\pi/N$ is the fixed point.

\section{Solution of the two channel anisotropic model}

In this section we discuss the two channel Kondo model in more detail,
focusing on the effects of channel anisotropy.
The starting hamiltonian is
\begin{equation}
H_K =
\sum_{a=1}^2 \sum_{\sigma}
H_0(\psi^{\phantom{\dagger}}_{a\sigma},
\psi^\dagger_{a\sigma}) +
\sum_{a=1}^2 \left\{ J_{z a} S^z \sigma_a^z({\bf 0})
+ \frac{J_{\perp a}}{2} \left[ S^+ \sigma_a^-({\bf 0})
+ S^- \sigma_a^+({\bf 0}) \right] \right\}.
\label{ham}
\end{equation}

If the exchange couplings are channel symmetric $J_{z1}=J_{z2}$
and $J_{\perp 1}=J_{\perp 2}$, it is known\cite{N&B} that the hamiltonian
(\ref{ham}) flows towards a non trivial fixed point.
At this fixed point the model exhibits non Fermi liquid behavior, namely
the impurity susceptibility $\chi_{imp}$ and the specific heat
over temperature $C_{V,imp}/T$ diverge at low temperatures as $\ln(1/T)$,
and the zero temperature entropy is finite and equal to $\ln(2)/2$,
as if half of the impurity spin degrees of freedom were decoupled from
the conduction electrons. Physically this occurs because two (and more)
channels tend to overscreen the impurity spin, so that the complete
screening characteristic of the single channel Kondo model can not
take place, thus leaving a ground state degeneracy.
In the case of a finite channel anisotropy, the system will
always choose the channel with the strongest exchange to screen the
impurity spin, and the usual Fermi liquid behavior of the single channel
model will finally take place at zero temperature. The corresponding
RG flow diagram is sketched in Fig.3.

As we have shown in the previous sections, this hamiltonian (\ref{ham})
can be mapped onto the following resonant level hamiltonian
\bea
H_{RL}&=&
H_0(\Psi,\Psi^\dagger) + H_0(\Psi_s^{\phantom{\dagger}},\Psi_s^\dagger)
+
\frac{J_{\perp 1}}{2}\sqrt{\nu_0\xi_0}
\left[d^\dagger \Psi({\bf 0}) + \Psi^\dagger({\bf 0}) d\right]\nonumber\\
&+&\frac{J_{\perp 2}}{2}\sqrt{\nu_0\xi_0}
\left[d^\dagger \Psi^\dagger({\bf 0}) + \Psi({\bf 0}) d\right]
+ \frac{W}{2}\left[
\Psi^\dagger({\bf 0})\Psi({\bf 0})
-\Psi({\bf 0})\Psi^\dagger({\bf 0})\right]
\left(d^\dagger d - \frac{1}{2}\right)\nonumber\\
&+& \frac{V}{2}\left[
\Psi_s^\dagger({\bf 0})\Psi_s^{\phantom{\dagger}}({\bf 0})
-\Psi_s^{\phantom{\dagger}}({\bf 0})\Psi_s^\dagger({\bf 0})\right]
\left( d^\dagger d -\frac{1}{2} \right) +
\Delta E  \left( d^\dagger d -\frac{1}{2} \right)
\label{full-hamres}
\eea
where the interaction potentials are related to
the longitudinal exchange couplings via
[see Eqs.(\ref{delta})-(\ref{V})-(\ref{W})]

\bea
W&=&\frac{1}{2} \frac{\displaystyle
J_{z 1} - J_{z 2} }
{\displaystyle
1 + \pi^2\nu_0^2 J_{z1}J_{z2}/16 },\\
V&=&\frac{8}{\pi^2\nu_0^2}\frac{\displaystyle
1 - \pi^2\nu_0^2 J_{z1}J_{z2}/16 }
{\displaystyle
J_{z1} + J_{z2} },
\eea
and (assuming equal impurity and conduction electron Land\'e factors
$g_i=g_c=g$)
\[
\Delta E = - \frac{2 g \mu_B}{\pi} {\rm tan}^{-1}
\left(\frac{\pi\nu_0 V}{2}\right) B \equiv -g\mu_B \Lambda(V) B.
\]

In the case of symmetric exchange couplings, the resonant
level model (\ref{full-hamres}) reduces to the hamiltonian
(\ref{EKham}) originally considered by Emery and Kivelson.
Then the combination $d^\dagger + d$ is
decoupled from the conduction electrons (hence the
ground state degeneracy and the non Fermi liquid behavior).
A finite channel anisotropic transverse exchange couples
this combination to conduction electrons and moves the system away
from the non Fermi liquid fixed point towards the
Fermi liquid single channel fixed point\cite{Sacramento}.
The smaller is the
anisotropy, the lower is the cross-over temperature.
In what follows we analytically study this cross-over in the
solvable limit $V=W=0$.

Since the total number of fermions is not conserved by the hamiltonian,
there are anomalous Green functions. In the Nambu
representation
\[
D=\left(
\begin{array}{c}
d \\
d^\dagger \\
\end{array}
\right),
\]
the impurity Green function
\[
\hat{G}_d(t)= - i\langle T\left( D(t)D^\dagger(0)\right)\rangle
\]
is a $2\times 2$ matrix. Its Fourier transform can easily be
evaluated. For $\omega$ much smaller than the bandwidth, we find:
\begin{equation}
\hat{G}_d(\omega) = \frac{1}{2}\,
\frac{\hat{\tau}_0 - \hat{\tau}_x}{\omega + i \Gamma {\rm sign}\omega}
+ \frac{1}{2}\,
\frac{\hat{\tau}_0 + \hat{\tau}_x}
{\omega + i \gamma {\rm sign}\omega}\; ,
\end{equation}
where the resonance widths are defined by
\begin{eqnarray*}
\Gamma &=& \frac{\pi}{4}\nu_0^2\xi_0\left(J_{\perp 1}+J_{\perp 2}\right)^2,\\
\gamma &=& \frac{\pi}{4}\nu_0^2\xi_0\left(J_{\perp 1}- J_{\perp 2}\right)^2,
\end{eqnarray*}
$\hat{\tau}_i$ being
the Pauli matrices, and $\hat{\tau}_0$ the unit matrix. The
impurity spectral function is
\[
\hat{A}(\omega)= \frac{1}{2}(\hat{\tau}_0 - \hat{\tau}_x)
\frac{\Gamma}{\omega^2 + \Gamma^2} +
\frac{1}{2}(\hat{\tau}_0 + \hat{\tau}_x)
\frac{\gamma}{\omega^2 + \gamma^2}\;,
\]
and it is therefore equally shared by two lorentzians with different
widths $\Gamma$ and $\gamma$. In the channel
isotropic case $\gamma\to 0$, one of the two lorentzians
tends to $\delta(\omega)$, representing the impurity
degree of freedom which is decoupled from the conduction band
in this particular limit\cite{E&K}.

The impurity contribution to the free energy can be calculated in a standard
way
by integration over the coupling constant.
The result is
\begin{equation}
F(T)=F_0(T) + \int \frac{d\omega }{2\pi} f(\omega)
\left[ {\rm tan}^{-1} \left(\frac{\Gamma}{\omega}\right)
+ {\rm tan}^{-1} \left(\frac{\gamma}{\omega}\right)
\right]\;,
\label{free-energy}
\end{equation}
where $F_0(T)$ is the free energy
in absence of coupling between the impurity and
conduction electrons, $f(\omega)$ is the Fermi distribution function
and the integral should be limited to the conduction
bandwidth. The entropy can be calculated by $S(T)=-\partial F(T)/\partial T$.
By defining the function
\be
\bar{S}(z)= \frac{1}{2\pi z}\left[
\psi\left( \frac{1}{2} + \frac{1}{2\pi z} \right) - 1 \right]
- {\rm ln}\Gamma\left(\frac{1}{2} + \frac{1}{2\pi z}
\right) +\frac{1}{2}\ln\pi\;,
\label{Sbar}
\ee
where $\psi(z)$ is the psi-function and $\Gamma(z)$ is the
gamma-function, the entropy turns out to be
\be
S(T)={\rm ln}(2) + \bar{S}\left(\frac{T}{\Gamma}\right) +
\bar{S}\left(\frac{T}{\gamma}\right)
= \left\{
\begin{array}{lr}
\displaystyle
\frac{\pi T}{6}\left( \frac{1}{\Gamma} + \frac{1}{\gamma}
\right)\;, & T\ll \gamma \\
\displaystyle
\ln \sqrt{2}\;, & \gamma \ll T \ll \Gamma \\
\displaystyle
\ln 2 - \frac{\Gamma + \gamma}{2\pi T}\;, &
T\gg \Gamma \\
\end{array}
\right.
\label{AA}
\ee
the last equality being valid for $\gamma\ll \Gamma$.
$S(T)$ is shown in Fig.4.
We see that $S(0)=0$, as expected since no degeneracy
is left, but there is a region of
temperatures (the wider the smaller $\gamma$ is)
where the entropy is close to that of the symmetric
two channel model.

Another quantity of physical interest
is the longitudinal impurity susceptibility. As we know from the
above analysis, exactly on the Emery-Kivelson
line, $\chi^{zz}_{imp}=0$ and one has to consider deviations
from this line (i.e. $V\not=0$) in order to account for a finite impurity
susceptibility\cite{others}.
The resulting susceptibility is
\begin{eqnarray*}
&&\chi^{zz}_{imp} = \left[g\mu_B \Lambda(V) \right]^2
\int_0^\beta d\tau \langle T\left( S^z(\tau) S^z(0) \right) \rangle=\\
&&
\left[g\mu_B \Lambda(V) \right]^2
\frac{1}{\pi(\Gamma-\gamma)}
\left[
\psi\left( \frac{1}{2} + \frac{\Gamma}{2\pi T} \right) -
\psi\left( \frac{1}{2} + \frac{\gamma}{2\pi T} \right)
\right] .
\end{eqnarray*}
In the case $\gamma \ll \Gamma$ the susceptibility shows the
same kind of cross-over behavior as the entropy:
\begin{equation}
\chi^{zz}_{imp} =
\left[g\mu_B \Lambda(V) \right]^2 \cdot
\left\{
\begin{array}{ll}
\displaystyle
\frac{1}{\pi(\Gamma-\gamma)}
\ln\left(\frac{\Gamma}{\gamma}\right)\;, & T\ll \gamma \\
\displaystyle
\frac{1}{\pi(\Gamma-\gamma)}
\ln\left(\frac{\Gamma}{T}\right)
\;, & \gamma \ll T \ll \Gamma \\
\displaystyle
\frac{1}{4 T}\;, &
T\gg \Gamma \\
\end{array}
\label{BB}
\right.
\end{equation}
As expected the magnetic susceptibility saturates at low
temperature, although at intermediate temperatures it shows
the logarithmic behavior of the two-channel Kondo model.

It follows from (\ref{AA}) and (\ref{BB}) that the Wilson ration $R_W$ is
{\em not} universal: it depends on the amount of anisotropy, $\gamma/\Gamma$.
Such a conclusion is obvious in the limit of small anisotropy, when the energy
scales are
well separated. Then the residual entropy $\ln\sqrt{2}$ must be quenched in a
temperature range
$\sim \gamma$, implying $C_{V, imp} \sim T/\gamma$, while the susceptibility
$\chi_{imp}$
just rounds off logarithmic singularity, $\chi_{imp}\sim \ln (\Gamma/\gamma)$:
the Wilson ratio is very {\em small}. Such a lack of universality is also
apparent in the
phenomenological, Fermi liquid description of the low temperature limit, $T\ll
\gamma$.
Then the impurity is quenched into a singlet, and the residual conduction
electron phase
shift in the channel $(m,\sigma)$ may be expanded as
\[
\delta_{m\sigma}(\epsilon)=\delta_{m0}+\alpha_m \epsilon +\psi_m\delta
n_{m,-\sigma}+
\sum_{m'\neq m} \phi^{m'\sigma'}_{m\sigma}\delta n_{m'\sigma'}\;,
\]
where $\delta n_{m'\sigma'}$ is the change in the occupation measured from the
ground state.
Universality implies that $\delta_{m\sigma}(\epsilon)$ is invariant
\begin{itemize}
\item[({\em i})] If the chemical potential of the other channel is changed
(there is no channel flip)
\item[({\em ii})] If $\epsilon$ and the chemical potential are changed by the
same
amount (the Kondo singularity is attached to the Fermi level)
\end{itemize}
Hence in our two channel case
\[
\left\{
\begin{array}{ll}
\delta_{1\sigma} & =\delta_{10}+\alpha_1\left[\epsilon - \frac{
\displaystyle \delta n_{1-\sigma}}{\displaystyle \nu_s}\right]
+\theta_1\sigma\sigma'\delta n_{2\sigma'} \\
\delta_{2\sigma} & =\delta_{20}+\alpha_2\left[\epsilon - \frac{
\displaystyle \delta n_{2-\sigma}}{\displaystyle \nu_s}\right]
+\theta_2\sigma\sigma'\delta n_{1\sigma'} \\
\end{array}
\right.
\]
(the cross terms $\theta_1, \theta_2$ are equal in the electron-hole symmetric
case $\delta_{10}=\delta_{20}
=\pi/2$.) $\nu_s$ is the one channel density of $s$-states at the Fermi level.
It is then
straightforward to extend the analysis of Ref.\onlinecite{N&B}: the resulting
impurity corrections are
\[
\frac{C_{V,imp}}{C_V}= \frac{\alpha_1+\alpha_2}{\pi \nu_s}\;,\;\;\;
\frac{\chi_{imp}}{\chi}= \frac{2(\alpha_1+\alpha_2)}{\pi
\nu_s}+\frac{\theta_1+\theta_2}{2\pi}
\]
Due to the channel interaction $\theta$ the Wilson ratio $R_W= C_V \chi_{imp}/
\chi C_{V, imp}$
departs from the single channel value $2$. Put another way, one has a line of
fixed points rather than
a unique one. If $"1"$ is the dominant screening channel, $J_1$ goes to
infinity while $J_2$ may
evolve towards any arbitrary value: once the spin $S$ is screened, $J_2$ no
longer scales.
This arbitrariness is reflected in the Wilson ratio.

To our knowledge, the two channel Kondo model is most convincingly
realized by two level systems in
metal alloys\cite{TLS}. This has recently been
experimentally confirmed thanks to the development
of point contact spectroscopy\cite{Ralph}.
In these systems, the role
of the spin is played by some orbital degree of freedom, while
the physical spin plays the role of the channel index. Thus
the model is by construction channel isotropic.
However,
an external magnetic field breaks the channel symmetry
and generates an effective channel anisotropy proportional to the
curvature of the conduction electron band times the magnetic field $B$.
In this case, the coupling to the magnetic field is described by the
following term in the hamiltonian:
\[
H_B=-\frac{g_c \mu_B }{2} B \int d{\bf x}
\sum_\sigma \left[
\psi_{1\sigma}^\dagger({\bf x})
\psi_{1\sigma}^{\phantom{\dagger}}({\bf x})
- \Psi_{2\sigma}^\dagger({\bf x})
\Psi_{2\sigma}^{\phantom{\dagger}}({\bf x})\right]
\]
where $\sigma$ is now the pseudo-spin index, and the channel indices
$1$ and $2$ correspond to the physical spin up and down respectively.
The magnetic field shifts the Fermi level of channel 1 with respect to
that of channel 2.
Electron-hole symmetry is thereby broken within each channel.
It follows that the magnetic field induces a phase shift anisotropy,
$\delta_1 -\delta_2 \propto B$,
reflected into a finite $W \propto B$.
Such a correction comes both from the correction to the Fermi level density of
states
\[
\nu_1 - \nu_2 = g_c \mu_B \nu_0' B\;, \;\;\;
\nu_0'=\left. \frac{\partial \nu}{\partial \epsilon}
\right|_{\epsilon=\epsilon_F}
\]
and from the Zeeman shift of band edges. The change of the density of states
also modifies
the pseudospin flip amplitudes in the equivalent
resonant level model:
\[
\frac{J_{\perp}}{2}\sqrt{\nu_1\xi_0}
\left[d^\dagger \Psi({\bf 0}) + \Psi^\dagger({\bf 0}) d\right]
+\frac{J_{\perp }}{2}\sqrt{\nu_2\xi_0}
\left[d^\dagger \Psi^\dagger({\bf 0}) + \Psi({\bf 0}) d\right] .
\]
The hybridization anisotropy is equivalent to a finite $\gamma$
\be
\gamma=\frac{\pi}{16} (g\mu_B)^2 \xi_0
\frac{(\nu_0')^2}{\nu_0} J_\perp^2 B^2 .
\label{gamma-B}
\ee
Thus, $B$ causes the cross-over to a Fermi liquid behavior at
low temperature as observed in Ref.\onlinecite{Ralph}.
As to the physical magnetic susceptibility, it is
related to the first derivative of the free
energy (\ref{free-energy}) with respect to $\gamma$.
The low temperature (low magnetic field) behavior of
the susceptibility is given by:
\[
\chi_{imp}=
\frac{(g\mu_B \nu_0')^2}{16 \nu_0} \xi_0 J_\perp^2
\ln\left[ {\rm min}\left(\frac{1}{T},\frac{1}{B^2}\right)\right].
\]
(One can show, that finite $W\propto B$ does not contribute to
the log-divergent part of the susceptibility.)

Very recently,
the channel anisotropic (but spin isotropic) Kondo model has been
solved using Bethe ansatz methods by N. Andrei and A. Jerez\cite{andrei}.
Their conclusions are qualitatively similar ours.

We are thankful to N. Andrei for helpful discussions.

\begin{figure}
\caption{Time-dependent potential seen by a spin up conduction electron.
The impurity spin flips from $\downarrow$ to
$\uparrow$ at times $t_i$ and vice versa at times $t'_j$.}
\label{Fig1}
\end{figure}

\begin{figure}
\caption{Time-dependent local potential felt by an electron
in the effective resonant level model. While the
same annihilation operator $\Psi$ is involved in the
two types of flips, the potentials are opposite.
The flips at times $t'_{i1}$ and $t_{j2}$ are the
hermitian conjugates.}
\label{Fig2}
\end{figure}

\begin{figure}
\caption{Qualitative $(J_1,J_2)$ RG flow diagram for the anisotropic
two channel Kondo model.}
\label{Fig3}
\end{figure}
\begin{figure}
\caption{Entropy $S(T)$ for various values of the anisotropy
$\lambda^2 = \gamma/\Gamma$: from the top $\lambda=0,0.1,0.5,1$.}
\label{Fig4}
\end{figure}

\end{document}